\documentclass[10pt,aps,prd,twocolumn,superscriptaddress,nofootinbib,floatfix]{revtex4-2}

\usepackage{placeins}
\usepackage{amsmath,amssymb}
\usepackage{graphicx}
\usepackage{xcolor}
\usepackage{multirow}
\usepackage{array}
\usepackage{ragged2e}
\usepackage[normalem]{ulem}

\usepackage[colorlinks=true,linkcolor=red,citecolor=green,urlcolor=blue]{hyperref}

\newlength{\colA}
\newlength{\colB}
\newlength{\colC}
\newlength{\colD}

\begin{document}

\title{Leptonic CP Conservation and the Quark CP Phase from Octonionic Flavor Structure}

\author{Bishnu Gupta Teli}
\affiliation{Department of Physics, Indian Institute of Technology Madras, Chennai 600036, India}

\author{Tejinder P. Singh}
\affiliation{Tata Institute of Fundamental Research, Homi Bhabha Road, Mumbai 400005, India}

\date{\today}

\begin{abstract}
One generation of standard-model fermions can be realized on the complexified octonions through the Clifford algebra $\mathcal{C}l(6)$; the octonionic unification programme extends this to three generations, with generation transport implemented by $G_2$ automorphisms or by rotors built from the ladder operators. We prove a localization theorem for the CP-violating phases of this structure, using only the $\mathcal{C}l(6)$ construction and the stated three-generation representatives, independently of the wider programme. For quarks, the first-to-second generation step is the occupation flip of one ladder mode, with the up and down species coupling to conjugate ladder directions; a conjugation theorem forces $A_d=A_u^*$ for every real transport, and the most general rung-generated rotor yields the exact one-parameter law $\phi_{12}=-2\chi$: the $(1,2)$ transport phase is twice one Yukawa orientation angle. The programme's geometric rotor sits exactly at the quadrature-balanced point $|\phi_{12}|=\pi/2$; the companion analysis reproduces the Cabibbo \emph{magnitude} $|V_{us}|$ with a single real tilt, leaving the rung near quadrature, but it does not extract a CKM CP phase, so the quark Dirac phase is fixed only once the underlying Yukawa orientation is computed. For leptons we prove a reality theorem: every charged-lepton and every neutrino transport amplitude is exactly real for every $G_2$ automorphism and every rotor that does not mix the identity line $\mathbb C\cdot1$ with the lepton--flavor plane $\mathrm{span}(e_7,e_5,e_2)$ a class that contains the entire quark-rung family--and identity--flavor mixing across that plane is the unique possible source of a leptonic phase. Leptonic Dirac CP is then conserved, $J_\ell=0$ and $\delta^\ell_{\rm CP}\in\{0,\pi\}$, unless the Higgs bridge mixes the identity line with that plane, the sole loophole, which we make precise but do not compute. Under a stated minimality assumption, the leading-order neutrino Jordan spectrum $(-\delta_\nu,0,+\delta_\nu)$ then selects the inverted mass ordering with opposite Majorana parities, giving $m_{\beta\beta}\simeq19$~meV and $\Sigma m_\nu\simeq0.10$~eV. The minimal lepton sector thus yields two sharply falsifiable predictions: the inverted ordering (JUNO) and $m_{\beta\beta}\simeq19$~meV (next-generation $0\nu\beta\beta$) together with a conditional expectation of leptonic CP conservation whose violation would not refute the framework but instead localize the missing bridge term (DUNE, Hyper-Kamiokande). We make the present confrontation explicit and quantify the lean, which currently runs against the lepton package on each axis.
\end{abstract}

\maketitle

\paragraph*{Introduction.---}
The octonionic unification programme~\cite{KaushikVaibhavSingh2022, Adler2004book, GunaydinGursey1973, Furey:2015yxg, furey_2018_plb, VaibhavSingh2023, Baez2002, schafer_nonassociative} seeks to obtain the standard-model gauge structure, three fermion generations, and the fermion mass ratios from the complexified octonions and the exceptional Jordan algebra $J_3(\mathbb O_{\mathbb C})$, within a pre-quantum trace dynamics. Square-root mass ratios for the charged fermions were derived from Jordan eigenvalue spectra in~\cite{Bhatt:2021adg} and rederived with sharper methodology in~\cite{Teli:2026jgr, Singh:2025xxv}; a falsification-oriented catalogue of the programme's experimental predictions appears in~\cite{Singh:2026cat}.

In this Letter, we ask a question the programme makes well-posed: where, in this algebra, can a CP-violating phase live? We answer it exactly, at the level of flavor transport. The result is asymmetric: the quark sector supports exactly one transport phase, which we derive in closed form, whereas the lepton sector supports none within the entire transport class employed by the programme. This is a theorem whose boundary we characterize exactly: identity--flavor mixing is the unique possible source of a leptonic phase, and within the rotor and channel transports of the construction it is also sufficient. Flavor mixing in the Standard Model is encoded in the CKM matrix for quarks~\cite{Cabibbo1963, KobayashiMaskawa1973} and the PMNS matrix for leptons~\cite{Pontecorvo1968, MakiNakagawaSakata1962}; CP-odd information is convention-independently captured by Jarlskog invariants~\cite{Jarlskog1985}, and current data are summarized in~\cite{ParticleDataGroup:2024cfk, Esteban:2024fit, NuFIT6.1}.

The analysis is self-contained at the level of Eqs.~\eqref{eq:ladder}--\eqref{eq:quark-reps}: it requires only the $\mathcal{C}l(6)$ ladder construction~\cite{Furey:2015yxg, furey_2018_plb} extended to three generations by the stated flavor representatives, and is logically independent of the wider trace-dynamics and Jordan-spectral programme, which enters only downstream through the companion's CKM magnitude fit~\cite{Singh:2025xxv}, the neutrino spectrum~\eqref{eq:sym-lam}, and the Higgs-bridge reading of the loophole. Throughout, we tag claims as derived ([D]), programme input ([P]), or open ([O]). We keep three statuses cleanly separate: the quark rung law is an exact transport theorem [D]; the lepton statement is exact once the final effective bridge is a safe real-linear transport [D, conditional]; and the neutrino mass-ordering and $0\nu\beta\beta$ commitments require an additional minimality assumption on the neutral-sector spectrum [P].

\begin{figure}[htbp]
\centering
\includegraphics[width=0.34\textwidth]{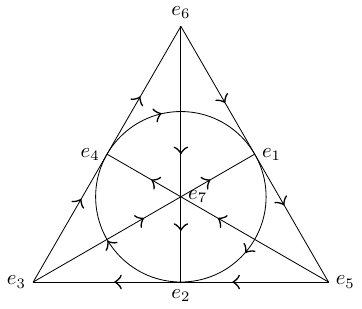}
\caption{Octonionic Multiplication Rules}
\label{fig:fano}
\end{figure}

\paragraph*{One Generation from $\mathcal{C}l(6)$.---}
We use the standard octonion units $e_1,\ldots,e_7$ multiplying along the seven directed Fano lines $(124),(235),(346),(615),(371),(574),(672)$ (so $e_1e_2=e_4$ cyclically, $e_ae_b=-e_be_a$, $e_a^2=-1$), shown in Fig.~\ref{fig:fano}, with an external complex unit $i$ commuting with all $e_a$~[P].
States are elements of $\mathbb C\otimes\mathbb O$, operators are composed left multiplications, and the Hermitian inner product is $\langle x|y\rangle=[\bar x^{\,*}y]_0$: octonionic conjugation and complex conjugation on $x$, multiply, keep the scalar part.

A maximal totally isotropic set of ladder operators is~[P]
\begin{align}\label{eq:ladder}
    \alpha_1=\frac{-\overleftarrow{e_5}+i \overleftarrow{e_4}}{2},
    \quad \alpha_2=\frac{-\overleftarrow{e_3}+i \overleftarrow{e_1}}{2},
    \quad \alpha_3=\frac{-\overleftarrow{e_6}+i \overleftarrow{e_2}}{2},
\end{align}
satisfying $\{\alpha_j,\alpha_k\}=0$, $\{\alpha_j,\alpha_k^\dagger\}=\delta_{jk}$, with idempotent $\bar V_\nu=\omega\omega^\dagger=(1+ie_7)/2$. Acting with $\alpha_k^\dagger$ on $\bar V_\nu$ generates one generation~\cite{Furey:2015yxg, furey_2018_plb}: the anti-down triplet $\alpha_k^\dagger\bar V_\nu$, the up triplet $\alpha_j^\dagger\alpha_k^\dagger\bar V_\nu$, and the positron $\alpha_3^\dagger\alpha_2^\dagger\alpha_1^\dagger\bar V_\nu\propto(i+e_7)$. The number operator $N=\sum_k\alpha_k^\dagger\alpha_k$ gives the electric charge $Q=N/3$.

The flavor representatives of the three generations used below are~[P]
\begin{align}\label{eq:lepton-reps}
    \nu_{L,k}\propto i\,e_{a_k},\qquad e^+_{L,k}\propto\pm(i\cdot1+e_{a_k}),
\end{align}
with $(a_1,a_2,a_3)=(7,5,2)$, and, for the first two quark generations,
\begin{align}\label{eq:quark-reps}
\begin{aligned}
    u_1&\propto e_4+ie_5, & \bar d_1&\propto e_5+ie_4,\\
    u_2&\propto e_6+ie_2, & \bar d_2&\propto e_2+ie_6.
\end{aligned}
\end{align}
Generation labels are assigned by the relabeling map $e_7\to e_5\to e_2\to e_3\to e_4\to e_6$ ($e_1$ fixed) on the coefficient space bookkeeping, not an automorphism; nothing below depends on which legitimate transport implements it. The quark--lepton split in available transports is itself structural~[D]: an explicit order-three automorphism $\Gamma\in\mathrm{Stab}_{G_2}(e_1)$ ($e_7\to e_5\to e_2\to e_7$; $e_3\to-e_6$, $e_6\to e_4$, $e_4\to-e_3$) realizes the colorless triples of Eq.~\eqref{eq:lepton-reps} as exact orbits ($\Gamma\,\nu_{L,g}=\nu_{L,g+1}$, $\Gamma\,e^+_{L,g}=\pm e^+_{L,g+1}$), but no automorphism extends this to the quarks the colorless cycle forces $e_4\mapsto e_2e_5=-e_3$, never $e_4\mapsto e_6$ and a conserved-invariant no-go in the companion~\cite{Singh:2025xxv} excludes the colored triples as $SU(3)_F$ orbits altogether. Quark generation transport is therefore necessarily of ladder type (the occupation flip below), while the lepton generation map is itself an automorphism, landing inside the real transport class of the theorem below. In contrast,

\paragraph*{A Cautionary Phase.---}
The elementary lepton overlaps of Eq.~\eqref{eq:lepton-reps} are purely imaginary: $\langle e^+_k|\nu_k\rangle\propto-[(i+e_a)(ie_a)]_0=i$. A naive intertwiner built from raw overlaps therefore carries a factor $i$ on an entire row of $U_{\rm PMNS}$, tempting one to read off $|\delta^\ell_{\rm CP}|=\pi/2$. But a row-global $i$ is a one-sided diagonal rephasing, removable by redefining the charged-lepton fields, and contributes nothing to the rephasing-invariant Jarlskog determinant $J_\ell=\Im(U_{e1}U_{\mu2}U^*_{e2}U^*_{\mu1})$~\cite{Jarlskog1985}. This artifact invites a maximal-leptonic-phase reading; the theorem below shows why no admissible transport could produce a physical version of it. Physical CP violation requires an interference phase, sandwiched between two nontrivial rotations; the quark sector supplies exactly one, which we now derive.

\paragraph*{The Quark CP Phase from the Rung.---}
The quark states~\eqref{eq:quark-reps} are ladder monomials: $u_1=\alpha_3^\dagger\alpha_2^\dagger\bar V_\nu$, $u_2=\alpha_3^\dagger\bar V_\nu$, $\bar d_1=\alpha_1^\dagger\bar V_\nu$, $\bar d_2=\alpha_2^\dagger\alpha_1^\dagger\bar V_\nu$. Direct computation gives the exact relations~[D]
\begin{equation}\label{eq:flip}
    \alpha_2 u_1=-u_2,\quad \alpha_2^\dagger\bar d_1=+\bar d_2,\quad \alpha_2^\dagger u_1=\alpha_2\bar d_1=0.
\end{equation}
The first-to-second generation step is therefore the occupation flip of the single rung mode $\alpha_2$: the up species empties it, the down species fills it. This explains structurally why the $(1,2)$ block is special: the $(2,3)$ step is a two-mode process ($\langle u_3|\alpha_1^\dagger u_2\rangle\neq0\neq\langle u_3|\alpha_3 u_2\rangle$), consistent with the anomalous normalization of that block in~\cite{Singh:2025xxv}~[D].

\paragraph*{Conjugation Theorem.---}
Let $K$ denote external complex conjugation ($i\to-i$, octonion units fixed). From Eqs.~\eqref{eq:quark-reps}, $\bar d_k=i\,K(u_k)$ exactly~[D]. Let $U$ be any real transport: any left rotor $L_{\exp(\theta g)}$ with $g$ a real imaginary octonion, or any $G_2$ automorphism. Since $K$ is a ring map on $\mathbb C\otimes\mathbb O$ that commutes with real transports, and the phases $i$ cancel in the Hermitian inner product, one finds
\begin{equation}\label{eq:conjugation}
    A_d\equiv\langle\bar d_2|U|\bar d_1\rangle=\langle u_2|U|u_1\rangle^*\equiv A_u^*
\end{equation}
for every such $U$~[D]. Thus $|A_u|=|A_d|$ is forced (the root-sum rule of~\cite{Bhatt:2021adg, Singh:2025xxv} inherits symmetric magnitudes), and the relative phase of the two interfering amplitudes is $\phi_{12}=\arg(A_u/A_d)=2\arg A_u$: the entire Cabibbo phase is the doubled phase of a single amplitude, verified numerically over $300$ random real rotors and automorphisms.

\paragraph*{The Rung Rotor and the Law $\phi_{12}=-2\chi$.---}
The most general unitary generated by the rung mode alone is $U(\theta,\chi)=\exp[\theta(e^{i\chi}\alpha_2^\dagger-e^{-i\chi}\alpha_2)]$, where $e^{i\chi}$ is the phase of the rung coupling. Using Eq.~\eqref{eq:ladder} one finds the operator identity~[D]
\begin{equation}\label{eq:gchi}
    e^{i\chi}\alpha_2^\dagger-e^{-i\chi}\alpha_2=\cos\chi\,e_3-\sin\chi\,e_1\equiv g_\chi,
\end{equation}
a real unit direction in the $(e_1,e_3)$ plane, so that $U(\theta,\chi)=L_{\exp(\theta g_\chi)}$ is left multiplication by $\exp(\theta g_\chi)$. Acting on the states and using Eq.~\eqref{eq:flip},
\begin{equation}\label{eq:amps}
    A_u=N e^{-i\chi}\sin\theta,\qquad A_d=N e^{+i\chi}\sin\theta,
\end{equation}
with $N$ real, hence the exact one-parameter law~[D]
\begin{equation}\label{eq:law}
    \phi_{12}=-2\chi.
\end{equation}
At full angle, $U(\tfrac{\pi}{2},\chi)$ transports $u_1\to e^{-i\chi}u_2$ and $\bar d_1\to e^{+i\chi}\bar d_2$, realizing the generation relabeling with species-opposite phases.

Equation~\eqref{eq:law} closes the kinematics of the rung principle. The rung quadratures are the pinned directions $\alpha_2+\alpha_2^\dagger=ie_1$ and $\alpha_2^\dagger-\alpha_2=e_3$; $\chi$ is the orientation of the coupling in this plane, physical because the frame is pinned ($e_1$ is the unit stabilized by the flavor $SU(3)\subset G_2$), so $\chi$ cannot be rotated away. Pure-quadrature couplings are CP-conserving ($\chi=0\Rightarrow\phi_{12}=0$; $\chi=\pi/2\Rightarrow\phi_{12}=\pi$), and $|\phi_{12}|=\pi/2\iff\chi=\mp\pi/4$: maximal CP violation is precisely the signature of a quadrature-balanced coupling.

\paragraph*{Relation to the Companion CKM Fit.---}
The geometric rotor of the companion analysis~\cite{Singh:2025xxv} sits exactly at the balanced point: it is amplitude-equivalent to $\chi=-\pi/4$, with $A_u\propto(1+i)$, $A_d\propto(1-i)$ and $\phi_{12}=\pi/2$~[D]. The companion fixes the Cabibbo \emph{magnitude} $|V_{us}|$ by adding a single real tilt of the up leg, $\varepsilon\simeq-26.1^\circ$, determined by $|V_{us}|=0.22497$ and \emph{not} by CP data~[P]; this leaves the rung near-balanced, displaced from quadrature by only $\simeq13^\circ$,
\begin{equation}\label{eq:chieff}
    \chi_{\rm eff}=-\tfrac12\!\left(\tfrac{\pi}{2}+\varepsilon\right)\simeq-32^\circ .
\end{equation}
We stress what is \emph{not} claimed. The companion extracts only CKM moduli ($|V_{us}|,|V_{cb}|,|V_{ub}|,|V_{td}|/|V_{ts}|$) and the Wolfenstein magnitudes; it does \emph{not} compute the rephasing-invariant CKM phase, the Jarlskog invariant, or the unitarity-triangle angle, and it does \emph{not} identify the $(1,2)$-block phase with the standard $\delta^{\rm quark}_{\rm CP}$~[O]. Converting a block phase to the convention-independent $\delta_{\rm CP}$ requires the full $(1,3)$ and $(2,3)$ structure, which lies outside both the companion's magnitude fit and the present transport theorem; moreover the tilt that fixes $|V_{us}|$ was fixed to a magnitude, so no CP datum is reproduced here even in principle. The robust statement is therefore structural, not numerical: by Eq.~\eqref{eq:law} the entire available quark phase is one Yukawa orientation angle, sitting near quadrature; deriving that angle and hence any physical CP phase awaits the Higgs-bridge vacuum~[O]. Everything else, the conjugate-pair structure, equal magnitudes, the doubling, the law~\eqref{eq:law}, the special status of the $(1,2)$ block, is derived.

\paragraph*{The Lepton Reality Theorem.---}
We now show that no analogous phase exists for leptons, and we characterize exactly which transports could ever produce one. Let $\Pi_\ell\equiv\mathrm{span}(e_7,e_5,e_2)$ denote the   lepton--flavor plane, the octonionic directions carried by the representatives~\eqref{eq:lepton-reps}. Let $U$ be any transport that does not mix the identity line $\mathbb C\cdot1$ with $\Pi_\ell$, i.e. $[U(1)]_{e_b}=0=[U(e_a)]_0$ for $a,b\in\{7,5,2\}$. This class contains: every automorphism $U\in G_2$, which fixes $1$ and preserves $\Im\,\mathbb O$ (in particular the flavor $SU(3)=\mathrm{Stab}_{G_2}(e_1)$); every global rotor $L_{\exp(\theta g)}$ with $g\perp\Pi_\ell$, i.e. $g\in\mathrm{span}(e_1,e_3,e_4,e_6)$; every channel rotor whose generator is orthogonal to the channel it rotates; hence the entire quark-rung family $U(\theta,\chi)=L_{\exp(\theta g_\chi)}$ of Eq.~\eqref{eq:gchi} for every $\chi$, and its $|V_{us}|$-tilt deformation, whose generators also lie in $\mathrm{span}(e_1,e_3)$. Then~[D]
\begin{equation}\label{eq:reality}
    \langle\nu_{L,j}|U|\nu_{L,k}\rangle\in\mathbb R,\qquad \langle e_j|U|e_k\rangle\in\mathbb R,
\end{equation}
for all generations $j,k$ and all such $U$.

\paragraph*{Proof.} Every transport above is complex-linear with real matrix elements in the basis $\{1,e_1,\ldots,e_7\}$. For any such map $\operatorname{C}$, expanding the states in components gives $\langle\nu_j|\operatorname{C}|\nu_k\rangle=[\operatorname{C} e_{a_k}]_{e_{a_j}}\in\mathbb R$: neutrino amplitudes are real for every real-linear transport, without further hypothesis. For the charged pair, one finds the exact master formula
\begin{equation}\label{eq:master}
    \langle e_j|\operatorname{C}|e_k\rangle=[\operatorname{C}(1)]_0+[\operatorname{C} e_{a_k}]_{e_{a_j}}+i\big([\operatorname{C}(1)]_{e_{a_j}}-[\operatorname{C} e_{a_k}]_0\big),
\end{equation}
up to immaterial overall real signs: the imaginary part is precisely the identity--flavor mixing of $\operatorname{C}$ across $\Pi_\ell$. Automorphisms kill both terms because they fix $1$ and preserve $\Im\,\mathbb O$; for a global rotor $L_{\exp(\theta g)}$ the two terms are $\sin\theta\,g_{e_{a_j}}$ and $-\sin\theta\,g_{e_{a_k}}$, which vanish for all generation pairs if and only if $g$ has no component in $\Pi_\ell$; for a channel rotor they vanish if and only if the generator is orthogonal to the rotated lepton channel. $\blacksquare$

\paragraph*{Sharpness.---}
Equation~\eqref{eq:master} holds for every real-linear map, so identity--flavor mixing across $\Pi_\ell$ is the unique possible source of a charged-lepton phase. Within the rotor and channel transports the construction employs, nontrivial mixing is also sufficient: a global rotor feeds the diagonal block $j=k$ with $\Im\langle e_k|U|e_k\rangle\propto2\sin\theta\,g_{e_{a_k}}$ and a channel rotor contributes a single term, so reality of all blocks forces $g\perp\Pi_\ell$ there, mixing and phase are equivalent~[D]. (For general real-linear maps the criterion is necessary only: the real-orthogonal swap of $1$ with $(e_7+e_5+e_2)/\sqrt3$, fixing the complement, mixes maximally yet cancels in every block; such maps are neither automorphisms nor rotor chains.) The minimal failure within the ladder itself is the $\chi=0$ member of the $\alpha_1$-rung family, $e^{i\chi}\alpha_1^\dagger-e^{-i\chi}\alpha_1=\cos\chi\,e_5-\sin\chi\,e_4$: for the normalized second-generation charged lepton $x=(i+e_5)/\sqrt2$,
\begin{equation}\label{eq:counterexample}
    \langle x|L_{\exp(\theta e_5)}|x\rangle=e^{i\theta},
\end{equation}
maximally complex both terms of Eq.~\eqref{eq:master} fire, since the rotor moves $1$ into $e_5$ and $e_5$ into $-1$. The structural picture is exact: the lepton tower $(e_7,e_5,e_2)$ consists of the Majorana vacuum direction together with the position quadratures of the $\alpha_1$ and $\alpha_3$ rungs, so those two rung families break lepton reality at generic $\chi$, while the $\alpha_2$ (Cabibbo) plane $\mathrm{span}(e_1,e_3)$ is the unique rung plane disjoint from $\Pi_\ell$. The algebra confines Dirac CP violation to the coloured sector through this disjointness: the one phase the quark sector possesses lives in the only rung plane that cannot touch the lepton--flavor channels, so it cannot leak~[D]. This disjointness is, however, a property of the monomial assignment of Eq.~\eqref{eq:flip}: the $(1,2)$ step is carried by $\alpha_2$, the single rung whose plane avoids $\Pi_\ell$, and whether that assignment is forced by the one-generation $\mathcal{C}l(6)$ construction or is one of several admissible generation labelings is a question we leave to the companion analysis~[P]. Conversely, any extension recruiting the $\alpha_1$ or $\alpha_3$ rungs at generic quadrature angle forfeits the theorem for leptons and must be re-examined against Eq.~\eqref{eq:master}; within the transports presently employed, no such operation occurs.

Numerically: $\mathfrak g_2=\operatorname{Der}(\mathbb O)$ from Schafer derivations~\cite{schafer_nonassociative} has dimension $14$ and $e_1$-stabilizer dimension $8$; lepton imaginary parts vanish identically over the full safe class, quark amplitudes are generically complex under the same transports, and every failure mode above is reproduced exactly (ancillary files)~[D].

\paragraph*{Prediction and the Single Loophole.---}
Any lepton-sector intertwiner assembled from transports in the stated class is a real matrix; $U_{\rm PMNS}$ is then real up to removable diagonal phases (such as the global $i$ of the cautionary example above), and
\begin{equation}\label{eq:lepCP}
    J_\ell=0,\qquad \delta^\ell_{\rm CP}\in\{0,\pi\}
\end{equation}
conditional on all lepton-sector phases being of flavor-transport origin within the non-mixing class~[D, conditional]. The conditionality is precise, not decorative. The physical neutrino mass term is a Weinberg-type operator whose matrix elements involve the bosonized Higgs-bridge field $B_H$ of the underlying trace-dynamics construction~\cite{Singh:2026bridge}, at present specified structurally rather than as an operator with computable matrix elements~[P]. If the bridge acts through a real left-multiplication chain $\operatorname{C}$, Eq.~\eqref{eq:master} applies verbatim and exhibits the only escape route: in the $(1,2)$ block,
\begin{equation}\label{eq:loophole}
    \Im\,\langle e_2|\operatorname{C}|e_1\rangle=[\operatorname{C}(1)]_{e_5}-[\operatorname{C}(e_7)]_0,
\end{equation}
and similarly in the other blocks with $(e_5,e_7)$ replaced by the corresponding pair of   lepton--flavor directions. A leptonic Dirac phase requires the bridge to mix the identity direction with the lepton--flavor channels in either direction; a bridge that respects the transport structure cannot generate one~[D]. Within real bridge operators this is the sole escape; an intrinsically complex bridge, with matrix elements not representable through real transports, would be a logically separate route, which nothing in the programme presently motivates and which the same triage computation would expose~[O]. (The neutrino-side amplitudes are real for any real-linear bridge whatsoever; the entire vulnerability resides in the charged-lepton block.) Note the economy: the same bridge orientation question controls both sectors its quadrature angle in $\mathrm{span}(e_1,e_3)$, the unique rung plane disjoint from $\Pi_\ell$, is the quark CP phase via Eq.~\eqref{eq:law}; its identity--flavor mixing across $\Pi_\ell$ is the only possible source of a lepton phase via Eq.~\eqref{eq:loophole}, and nothing in the programme presently supplies it. Computing the matrix elements of $B_H$ between ladder states (``bridge triage'') would render both statements unconditional in one stroke~[O]. The size of any such phase calibrates this bridge directly: a near-conserving value would leave the real transport in control, with only a perturbative identity--flavor admixture, whereas a \emph{maximal} phase $\delta^\ell_{\rm CP}=\pm\pi/2$ would force that admixture to be $O(1)$ so that $B_H$, not the transport skeleton, governs the observed leptonic CP physics~[O]. Suggestively, $\pm\pi/2$ is the same quadrature-balanced value the quark rung realizes at $|\phi_{12}|=\pi/2$ (Eq.~\eqref{eq:law}): a bridge rotor seated at the analogous balance point would make maximal leptonic violation natural rather than anomalous, mirroring the coloured sector. Whether the localization dynamics selects that point is the open bridge computation, which a maximal measurement would fix from data rather than predict from theory~[O].

\paragraph*{Inverted Ordering and $0\nu\beta\beta$.---}
The neutrino sector adds an independent, sharper commitment. In the Jordan-spectral derivation of mass ratios~\cite{Bhatt:2021adg, Teli:2026jgr, Singh:2025xxv}, the right-handed Jordan eigenvalues of the neutrino family are
\begin{equation}\label{eq:sym-lam}
    (a_\nu,b_\nu,c_\nu)=(-\delta_\nu,0,+\delta_\nu),\qquad \delta_\nu^2=\tfrac34
\end{equation}
[P], i.e. one vanishing eigenvalue and two of equal magnitude and opposite sign. With the square-root-mass identification of~\cite{Bhatt:2021adg}, the leading-order mass pattern is $(m,m,0)$: a quasidegenerate pair and a massless state. Structurally, this is already the inverted pattern of two heavy states and one light, whereas the normal ordering requires one heavy and two light and is unreachable by small lifts of the symmetric spectrum. Making the assignment precise requires one further input, a minimality assumption~[P]: the subleading lifts that generate the solar splitting are perturbative, $|\text{corrections}|\ll\delta_\nu$, so they split the quasi-degenerate pair without reordering the spectrum. Under that assumption, assigning the pair to the solar doublet ($m_1\simeq m_2$) and the vanishing eigenvalue to $m_3$ reproduces $\Delta m^2_{21}\ll|\Delta m^2_{31}|$, whereas the normal ordering would require a non-perturbative rearrangement of the leading symmetric spectrum [D, given Eq.~\eqref{eq:sym-lam} and minimality]. We flag the second input deliberately: the general spectral fits of~\cite{Teli:2026jgr} and the companion texture with an unconstrained center lift~\cite{Singh:2025xxv} accommodate both orderings; it is the minimal, perturbative-lift reading of Eq.~\eqref{eq:sym-lam} that commits to inversion, which is precisely what makes it falsifiable: a normal-ordering determination falsifies the minimal reading, not a parameter choice.

The opposite signs in Eq.~\eqref{eq:sym-lam} are, for a real symmetric Majorana mass matrix, opposite Majorana CP parities of the quasidegenerate pair. With $m_3\simeq0$, $m_1\simeq m_2\simeq\sqrt{|\Delta m^2_{31}|}\simeq0.050$~eV and parities $\eta_1\eta_2=-1$, the effective $0\nu\beta\beta$ mass is driven to the lower edge of the inverted-ordering band:
\begin{equation}\label{eq:mbb}
    m_{\beta\beta}\simeq m\,c_{13}^2\,|\cos2\theta_{12}|\simeq19~\text{meV},
\end{equation}
using the global-fit angles~\cite{NuFIT6.1}, with $\Sigma m_\nu\simeq2m\simeq0.10$~eV [D, conditional on Eq.~\eqref{eq:sym-lam}]. We state the cosmological situation at its actual strength: the tightest minimal-$\Lambda$CDM combinations now lie below this sum DESI DR2 BAO with CMB gives $\Sigma m_\nu<0.064$~eV at $95\%$~\cite{DESI:2025dr2} so on that reading the spectrum~\eqref{eq:sym-lam} is already disfavored in minimal $\Lambda$CDM and is viable only in extended cosmologies: the bound relaxes to $\Sigma m_\nu<0.16$~eV in $w_0w_a$ dark energy~\cite{DESI:2025dr2} and under alternative Planck likelihoods and supernova choices~\cite{AllaliNotari2024}, and the same DESI combinations independently challenge $\Lambda$CDM itself. Invoking these extensions to accommodate $\Sigma m_\nu\simeq0.10$~eV is itself a model-dependent move; the cosmological verdict on Eq.~\eqref{eq:sym-lam} is therefore unsettled today and will sharpen rapidly~[O]. Meanwhile $m_{\beta\beta}\simeq19$~meV lies squarely in the discovery window of LEGEND-1000~\cite{LEGEND:2021preconceptual} and nEXO~\cite{nEXO:2021nexo}.

\paragraph*{Confrontation with Experiment and Falsification Outlook.---}
The framework now stands committed to three correlated statements that, \emph{given the single minimality assumption above, carry no further adjustable parameter}: inverted ordering; $m_{\beta\beta}\simeq19$~meV; and leptonic Dirac CP conservation, $\delta^\ell_{\rm CP}\in\{0,\pi\}$. The three are not equally robust. The first two are escapable only by altering the stated neutrino spectrum, Eq.~\eqref{eq:sym-lam}, and so are genuine falsifiable predictions; the third is escapable through the as-yet-uncomputed bridge term of Eq.~\eqref{eq:loophole}, so a measured leptonic phase would relocate that term rather than refute the construction. The present data keep each lepton-sector commitment alive separately while leaning against their conjunction, and we quantify the lean rather than soften it. On the quark side there is no comparable confrontation to report: the construction fixes the structural law $\phi_{12}=-2\chi$ and the Cabibbo \emph{magnitude} (through a tilt fit to $|V_{us}|$), but it does not yield a CKM CP phase, so there is no quark-sector CP number to test here the falsifiable content is the lepton package alone. Table~\ref{tab:expt} collects the confrontation.

\begin{table*}[htbp]
\centering

\setlength{\tabcolsep}{4pt}
\renewcommand{\arraystretch}{1.9}

\setlength{\colA}{0.235\textwidth}
\setlength{\colB}{0.135\textwidth}
\setlength{\colC}{0.345\textwidth}
\setlength{\colD}{0.175\textwidth}

\begin{tabular}{llll}
\hline\hline

\begin{minipage}[t]{\colA}
\textbf{Framework Statement}
\end{minipage}
&
\begin{minipage}[t]{\colB}
\textbf{Observable}
\end{minipage}
&
\begin{minipage}[t]{\colC}
\textbf{Present Status (The Lean)}
\end{minipage}
&
\begin{minipage}[t]{\colD}
\textbf{Decisive Test}
\end{minipage}
\\[0.4em]

\hline

\begin{minipage}[t]{\colA}
Quark CP Phase $=$ one Yukawa orientation, via $\phi_{12}=-2\chi$\\
{[D law; phase not predicted]}
\end{minipage}
&
\begin{minipage}[t]{\colB}
CKM $\delta_{\rm CP}$ / $J_q$
\end{minipage}
&
\begin{minipage}[t]{\colC}
\justifying\noindent companion reproduces $|V_{us}|$ by a tilt fit~\cite{Singh:2025xxv}; no $\delta_{\rm CP}$ extracted; the rung orientation $\chi$ is uncomputed.
\end{minipage}
&
\begin{minipage}[t]{\colD}
\justifying\noindent
awaits Higgs-bridge vacuum [Eq.~\eqref{eq:loophole}]
\end{minipage}
\\[0.8em]

\begin{minipage}[t]{\colA}
Leptonic Dirac CP conserved, $\delta^\ell_{\rm CP}\in\{0,\pi\}$\\
{[D, conditional; diagnostic]}
\end{minipage}
&
\begin{minipage}[t]{\colB}
$\nu/\bar\nu$ appearance asymmetry
\end{minipage}
&
\begin{minipage}[t]{\colC}
\justifying\noindent NuFit-6.0~\cite{Esteban:2024fit}: for inverted ordering, $\delta\simeq270^\circ$ favored over conservation at $>3.6\sigma$; for normal ordering, consistent within $1\sigma$. T2K$+$NOvA inverted $3\sigma$ interval $[-0.92\pi,-0.04\pi]$ excludes both CP-conserving values~\cite{T2K:2025wet}.
\end{minipage}
&
\begin{minipage}[t]{\colD}
\justifying\noindent DUNE~\cite{DUNEPhysicsTDR2020}, Hyper-K~\cite{HyperK:2025sensitivity} ($5\sigma$ over $>60\%$ of $\delta$)
\end{minipage}
\\[0.8em]

\begin{minipage}[t]{\colA}
Inverted mass ordering\\
{[P $+$ minimality]}
\end{minipage}
&
\begin{minipage}[t]{\colB}
reactor / atmospheric $\Delta m^2$
\end{minipage}
&
\begin{minipage}[t]{\colC}
\justifying\noindent
global fits mildly prefer normal ordering ($\Delta\chi^2\simeq6$ with SK atmospheric)~\cite{Esteban:2024fit}, though a combined fit to current accelerator data alone prefers inverted (resolving the NOvA-T2K tension)~\cite{Goswami:2026syn}; DESI$+$CMB $\Sigma m_\nu<0.064$~eV ($\Lambda$CDM)~\cite{DESI:2025dr2} disfavors inverted.
\end{minipage}
&
\begin{minipage}[t]{\colD}
\justifying\noindent JUNO~\cite{JUNO:2024jaw} ($\sim3\sigma$, ordering-only, $\sim6.5$~yr)
\end{minipage}
\\[0.8em]

\begin{minipage}[t]{\colA}
$m_{\beta\beta}\simeq19$~meV, $\Sigma m_\nu\simeq0.10$~eV\\
{[D, conditional on Eq.~\eqref{eq:sym-lam}]}
\end{minipage}
&
\begin{minipage}[t]{\colB}
$0\nu\beta\beta$; cosmology
\end{minipage}
&
\begin{minipage}[t]{\colC}
\justifying\noindent $m_{\beta\beta}$ within reach; $\Sigma m_\nu$ above tightest $\Lambda$CDM bound but relaxes to $<0.16$~eV in $w_0w_a$ and alternative likelihoods~\cite{DESI:2025dr2,AllaliNotari2024}.
\end{minipage}
&
\begin{minipage}[t]{\colD}
\justifying\noindent LEGEND-1000~\cite{LEGEND:2021preconceptual}, nEXO~\cite{nEXO:2021nexo}
\end{minipage}
\\[0.4em]

\hline\hline
\end{tabular}

\caption{Confrontation of the octonionic lepton CP sector with experiment. The quark sector yields the structural law $\phi_{12}=-2\chi$ but no CKM CP-phase prediction (the rung orientation is uncomputed; only $|V_{us}|$ is fit), so it carries no testable CP number here. The minimal lepton package is the falsifiable content: two predictions (ordering, $m_{\beta\beta}$) plus one diagnostic (leptonic CP, whose violation would locate the bridge term rather than refute the framework) which the present global fit disfavors at roughly the $3\sigma$ level on each axis, and which the next decade will settle.}
\label{tab:expt}

\end{table*}

Concretely: global fits mildly prefer normal ordering while allowing the inverted one; within the inverted ordering they favor near-maximal CP violation, $\delta_{\rm CP}\simeq270^\circ$, over CP conservation at more than $3.6\sigma$ in the published NuFit-6.0 analysis~\cite{Esteban:2024fit}, with~\cite{NuFIT6.1} the current release. The first joint T2K-NOvA analysis~\cite{T2K:2025wet} shows no strong preference for either ordering, but its inverted-ordering $3\sigma$ interval for $\delta_{\rm CP}$, $[-0.92\pi,-0.04\pi]$, excludes both CP-conserving values: if the ordering proves inverted, current long-baseline data already constitute evidence for leptonic CP violation. A recent combined GLoBES analysis of the current accelerator data~\cite{Goswami:2026syn} sharpens both sides of this ledger: with the mass ordering left free, NOvA-2024 and T2K-2020 jointly prefer the \emph{inverted} ordering largely because it removes the $\simeq2\sigma$ NOvA-T2K tension present under normal ordering so the long-baseline data alone lean, if anything, toward our ordering; but the same fit places that inverted solution at $\delta_{\rm CP}\simeq270^\circ$, with the competing CP-conserving minimum ($\delta_{\rm CP}\simeq180^\circ$) lying in the \emph{normal} ordering ($\Delta\chi^2\simeq1.6$). Each data-preferred solution therefore satisfies exactly one leg of our prediction, the ordering or the CP conservation, and it is precisely their conjunction that the present data do not select. The package predicted here, inverted ordering with conserved CP, thus stands against the present oscillation lean at roughly the $3\sigma$ level, and against the tightest minimal-$\Lambda$CDM cosmology independently. (T2K alone excluded both CP-conserving values at $95\%$~CL in 2020~\cite{T2K:2020}; NOvA remains consistent with conservation for normal ordering~\cite{NOvA:2022}.) The resolution will not wait long: JUNO will determine the ordering independently of $\delta_{\rm CP}$ and $\theta_{23}$, with $\simeq3\sigma$ median sensitivity in about $6.5$~years of nominal exposure~\cite{JUNO:2024jaw}; Hyper-Kamiokande can exclude CP conservation at $5\sigma$ for over $60\%$ of true $\delta_{\rm CP}$ values within its nominal ten-year exposure~\cite{HyperK:2025sensitivity}, with DUNE providing comparable, complementary reach~\cite{DUNEPhysicsTDR2020}. We consider this riskiness the appropriate standard: JUNO and the $0\nu\beta\beta$ programs can each falsify the minimal lepton sector outright, while a long-baseline determination of $\delta^\ell_{\rm CP}\notin\{0,\pi\}$ would instead pin the missing bridge term of Eq.~\eqref{eq:loophole}; jointly they will settle the package within the decade, or overturn the present lean and confirm a pattern inverted ordering, conserved leptonic CP, $m_{\beta\beta}$ at the band edge that no parameter of the framework can adjust.

On the quark side, the rung principle reduces the $(1,2)$ transport phase to one Yukawa orientation angle through the exact law $\phi_{12}=-2\chi$, with maximality the signature of quadrature balance; deriving $\chi$ from the Higgs-bridge vacuum which would for the first time turn the structural law into a genuine CKM CP-phase prediction and the bridge triage of Eq.~\eqref{eq:loophole}, are the two remaining computations, and they are the same computation. We close on a division of labour the reader should keep in view: the reality and rung theorems are self-contained but yield no number on their own, whereas the quantitative predictions of the neutrino spectrum~\eqref{eq:sym-lam}, and hence the ordering and $m_{\beta\beta}$, are inherited from the companion construction~\cite{Singh:2025xxv} and must ultimately be judged there.

\paragraph*{Numerical Verification.---}
The algebraic claims above were checked by an independent Python suite supplied as ancillary material: the Fano multiplication table, the composed left-multiplication Clifford operators, random exponentials of Schafer derivations of $\mathfrak g_2$~\cite{schafer_nonassociative}, the rung rotors, and the lepton master formula. The current run verifies the local rung law $\phi_{12}=-2\chi$, the conjugation relation $A_d=A_u^*$ over random real transports, the order-three flavor automorphism orbits, lepton reality under random $G_2$ automorphisms and under the entire $\alpha_2$ rung family, and explicit failure for scalar--flavor rotors and the $\alpha_1$/$\alpha_3$ counterexamples. The suite is not a substitute for the analytic proof; it fixes the nonassociative convention and checks that signs, adjoints, automorphisms, and rephasing artifacts match the formulas in the text. It does not test the neutrino mass-ordering, $m_{\beta\beta}$, or $\Sigma m_\nu$ statements, which are analytic consequences of Eq.~\eqref{eq:sym-lam} and the minimality assumption rather than transport identities.

\begin{acknowledgments}
The authors acknowledge extensive assistance from Claude Opus 4.8 (Anthropic) and GPT 5.5 Pro (OpenAI) in the derivations, the numerical verification, the external check of earlier drafts, and the preparation of this manuscript. The authors assume full intellectual responsibility for the content of the manuscript.

B.\,G.\,T. would like to thank Dawood Kothawala, Department of Physics, IIT Madras, for his guidance and support as co-supervisor, and for enabling this work to be partly carried out as part of the M.Sc. project. He also thanks the Inter-University Centre for Astronomy and Astrophysics (IUCAA), Pune, for hospitality and accommodation during a research visit. 

The verification suite (octonion engine, ladder identities, $G_2$/stabilizer construction, rung and transport scans, the counterexample~\eqref{eq:counterexample}, the tilted-rotor equivalence~\eqref{eq:chieff}, and the master-formula and cancellation checks, and the flavor-automorphism orbit checks) accompanies the submission as ancillary files.
\end{acknowledgments}

\bibliography{references}

@misc{Teli:2026jgr,
    author = "Teli, Bishnu Gupta and Singh, Tejinder Pal",
    title = "{Fermion Mass Hierarchies and the Exceptional Jordan Algebra}",
    eprint = "2605.24866",
    archivePrefix = "arXiv",
    primaryClass = "hep-ph",
    month = "5",
    year = "2026"
}

@phdthesis{Furey:2015yxg,
    author        = "Furey, C.",
    title         = "{Standard model physics from an algebra?}",
    school        = "University of Waterloo",
    year          = "2015",
    eprint        = "1611.09182",
    archivePrefix = "arXiv",
    primaryClass  = "hep-th"
}

@misc{NuFIT6.1,
  author       = {{NuFIT Collaboration}},
  title        = {{NuFIT 6.1 (2025)}},
  howpublished = {\url{http://www.nu-fit.org/?q=node/309}},
  year         = {2025},
  note         = {Global fit results for neutrino oscillation parameters}
}

@article{ParticleDataGroup:2024cfk,
    author = "Navas, S. and others",
    collaboration = "Particle Data Group",
    title = "{Review of particle physics}",
    doi = "10.1103/PhysRevD.110.030001",
    journal = "Phys. Rev. D",
    volume = "110",
    number = "3",
    pages = "030001",
    year = "2024"
}

@article{T2K:2025wet,
    author = "Abubakar, S. and others",
    collaboration = "T2K, NOvA",
    title = "{Joint neutrino oscillation analysis from the T2K and NOvA experiments}",
    eprint = "2510.19888",
    archivePrefix = "arXiv",
    primaryClass = "hep-ex",
    reportNumber = "FERMILAB-PUB-25-0132-PPD",
    doi = "10.1038/s41586-025-09599-3",
    journal = "Nature",
    volume = "646",
    number = "8086",
    pages = "818--824",
    year = "2025"
}

@article{JUNO:2024jaw,
    author = "Abusleme, Angel and others",
    collaboration = "JUNO",
    title = "{Potential to identify neutrino mass ordering with reactor antineutrinos at JUNO}",
    eprint = "2405.18008",
    archivePrefix = "arXiv",
    primaryClass = "hep-ex",
    doi = "10.1088/1674-1137/ad7f3e",
    journal = "Chin. Phys. C",
    volume = "49",
    number = "3",
    pages = "033104",
    year = "2025"
}

@misc{Singh:2025xxv,
    author = "Singh, Tejinder P.",
    title = "{Fermion mass ratios from the exceptional Jordan algebra}",
    eprint = "2508.10131",
    archivePrefix = "arXiv",
    primaryClass = "hep-ph",
    month = "8",
    year = "2025"
}

@article{Cabibbo1963,
    author = "Cabibbo, Nicola",
    title = "{Unitary Symmetry and Leptonic Decays}",
    doi = "10.1103/PhysRevLett.10.531",
    journal = "Phys. Rev. Lett.",
    volume = "10",
    pages = "531--533",
    year = "1963"
}

@article{KobayashiMaskawa1973,
    author = "Kobayashi, Makoto and Maskawa, Toshihide",
    title = "{CP Violation in the Renormalizable Theory of Weak Interaction}",
    reportNumber = "KUNS-242",
    doi = "10.1143/PTP.49.652",
    journal = "Prog. Theor. Phys.",
    volume = "49",
    pages = "652--657",
    year = "1973"
}

@article{MakiNakagawaSakata1962,
    author = "Maki, Ziro and Nakagawa, Masami and Sakata, Shoichi",
    title = "{Remarks on the unified model of elementary particles}",
    doi = "10.1143/PTP.28.870",
    journal = "Prog. Theor. Phys.",
    volume = "28",
    pages = "870--880",
    year = "1962"
}

@article{Pontecorvo1968,
    author = "Pontecorvo, B.",
    title = "{Neutrino Experiments and the Problem of Conservation of Leptonic Charge}",
    journal = "Sov. Phys. JETP",
    volume = "26",
    pages = "984--988",
    year = "1968"
}

@article{Jarlskog1985,
    author = "Jarlskog, C.",
    title = "{Commutator of the Quark Mass Matrices in the Standard Electroweak Model and a Measure of Maximal CP Nonconservation}",
    reportNumber = "USIP-85-14",
    doi = "10.1103/PhysRevLett.55.1039",
    journal = "Phys. Rev. Lett.",
    volume = "55",
    pages = "1039",
    year = "1985"
}

@article{Baez2002,
    author        = "Baez, John C.",
    title         = "{The Octonions}",
    journal       = "Bull. Am. Math. Soc.",
    volume        = "39",
    number        = "2",
    pages         = "145--205",
    year          = "2002",
    doi           = "10.1090/S0273-0979-01-00934-X",
    eprint        = "math/0105155",
    archivePrefix = "arXiv",
    note          = "[Erratum: Bull. Am. Math. Soc. 42, 213 (2005)]"
}

@misc{DUNEPhysicsTDR2020,
    author = "Abi, Babak and others",
    collaboration = "DUNE",
    title = "{Deep Underground Neutrino Experiment (DUNE), Far Detector Technical Design Report, Volume II: DUNE Physics}",
    eprint = "2002.03005",
    archivePrefix = "arXiv",
    primaryClass = "hep-ex",
    reportNumber = "FERMILAB-PUB-20-025-ND, FERMILAB-DESIGN-2020-02",
    doi = "10.2172/1599307",
    month = "2",
    year = "2020"
}

@article{GunaydinGursey1973,
    author = "Gunaydin, Murat and Gursey, Feza",
    title = "{Quark structure and octonions}",
    doi = "10.1063/1.1666240",
    journal = "J. Math. Phys.",
    volume = "14",
    pages = "1651--1667",
    year = "1973"
}

@article{Bhatt:2021adg,
    author = "Bhatt, Vivan and Mondal, Rajrupa and Vaibhav, Vatsalya and Singh, Tejinder P.",
    title = "{Majorana neutrinos, exceptional Jordan algebra, and mass ratios for charged fermions}",
    eprint = "2108.05787",
    archivePrefix = "arXiv",
    primaryClass = "hep-ph",
    doi = "10.1088/1361-6471/ac4c91",
    journal = "J. Phys. G",
    volume = "49",
    number = "4",
    pages = "045007",
    year = "2022"
}

@article{Esteban:2024fit,
    author = "Esteban, Ivan and Gonzalez-Garcia, M. C. and Maltoni, Michele and Martinez-Soler, Ivan and Pinheiro, Jo{\~a}o Paulo and Schwetz, Thomas",
    title = "{NuFit-6.0: updated global analysis of three-flavor neutrino oscillations}",
    eprint = "2410.05380",
    archivePrefix = "arXiv",
    primaryClass = "hep-ph",
    reportNumber = "IFT-UAM/CSIC-24-140, YITP-SB-2024-24, IPPP/24/64, IPPP/24/64, IFT-UAM/CSIC-24-140, YITP-SB-2024-24",
    doi = "10.1007/JHEP12(2024)216",
    journal = "JHEP",
    volume = "12",
    number = "12",
    pages = "216",
    year = "2024"
}

@misc{HyperK:2025sensitivity,
    author = "Abe, K. and others",
    collaboration = "Hyper-Kamiokande",
    title = "{Sensitivity of the Hyper-Kamiokande experiment to neutrino oscillation parameters using accelerator neutrinos}",
    eprint = "2505.15019",
    archivePrefix = "arXiv",
    primaryClass = "hep-ex",
    doi = "10.1140/epjc/s10052-025-14938-9",
    journal = "Eur. Phys. J. C",
    volume = "86",
    number = "2",
    pages = "170",
    year = "2026"
}

@misc{LEGEND:2021preconceptual,
    author = "Abgrall, N. and others",
    collaboration = "LEGEND",
    title = "{The Large Enriched Germanium Experiment for Neutrinoless $\beta\beta$ Decay}: {LEGEND-1000 Preconceptual Design Report}",
    eprint = "2107.11462",
    archivePrefix = "arXiv",
    primaryClass = "physics.ins-det",
    month = "7",
    year = "2021"
}

@article{nEXO:2021nexo,
    author = "Adhikari, G. and others",
    collaboration = "nEXO",
    title = "{nEXO: neutrinoless double beta decay search beyond 10$^{28}$ year half-life sensitivity}",
    eprint = "2106.16243",
    archivePrefix = "arXiv",
    primaryClass = "nucl-ex",
    doi = "10.1088/1361-6471/ac3631",
    journal = "J. Phys. G",
    volume = "49",
    number = "1",
    pages = "015104",
    year = "2022"
}

@book{schafer_nonassociative,
  title={An Introduction to Nonassociative Algebras},
  author={Schafer, R.D.},
  isbn={9780486688138},
  lccn={lc95022734},
  series={Dover Books on Mathematics},
  url={https://books.google.com.np/books?id=IsmW8-etBEwC},
  year={2017},
  publisher={Dover Publications}
}

@article{furey_2018_plb,
    author = "Furey, C.",
    title = "{$SU(3)_C\times SU(2)_L\times U(1)_Y\left( \times U(1)_X \right) $ as a symmetry of division algebraic ladder operators}",
    eprint = "1806.00612",
    archivePrefix = "arXiv",
    primaryClass = "hep-th",
    doi = "10.1140/epjc/s10052-018-5844-7",
    journal = "Eur. Phys. J. C",
    volume = "78",
    number = "5",
    pages = "375",
    year = "2018"
}

@misc{KaushikVaibhavSingh2022,
    author = "Kaushik, Priyank and Vaibhav, Vatsalya and Singh, Tejinder P.",
    title = "{An $E_8\otimes E_8$ unification of the standard model with pre-gravitation, on an octonion-valued twistor space}",
    eprint = "2206.06911",
    archivePrefix = "arXiv",
    primaryClass = "hep-ph",
    year = "2022"
}

@book{Adler2004book,
    author = "Adler, Stephen L.",
    title = "{Quantum Theory as an Emergent Phenomenon}",
    publisher = "Cambridge University Press",
    address = "Cambridge",
    year = "2004",
    doi = "10.1017/CBO9780511535277",
    isbn = "9780521831949"
}

@article{VaibhavSingh2023,
    author = "Vaibhav, Vatsalya and Singh, Tejinder P.",
    title = "{Left-Right Symmetric Fermions and Sterile Neutrinos from Complex Split Biquaternions and Bioctonions}",
    eprint = "2108.01858",
    archivePrefix = "arXiv",
    primaryClass = "hep-ph",
    journal = "Adv. Appl. Clifford Algebras",
    volume = "33",
    number = "3",
    pages = "32",
    year = "2023"
}

@misc{Singh:2026cat,
    author = "Singh, Tejinder P.",
    title = "{Experimental predictions of the $E_8\otimes E_8$ octonionic unification programme: a falsification-oriented catalogue}",
    eprint = "2604.06288",
    archivePrefix = "arXiv",
    primaryClass = "hep-ph",
    year = "2026"
}

@misc{Singh:2026bridge,
    author = "Singh, Tejinder P.",
    title = "{Towards deriving the Standard Model coupled to gravity from Generalized Trace Dynamics via the spectral action principle}",
    howpublished = "Preprints 2026, 2026051806",
    doi = "10.20944/preprints202605.1806.v1",
    year = "2026"
}

@article{DESI:2025dr2,
    author = "Abdul-Karim, M. and others",
    collaboration = "DESI",
    title = "{DESI DR2 Results II: Measurements of Baryon Acoustic Oscillations and Cosmological Constraints}",
    eprint = "2503.14738",
    archivePrefix = "arXiv",
    primaryClass = "astro-ph.CO",
    journal = "Phys. Rev. D",
    volume = "112",
    number = "8",
    pages = "083515",
    year = "2025"
}

@misc{AllaliNotari2024,
    author = "Allali, Itamar J. and Notari, Alessio",
    title = "{Neutrino mass bounds from DESI 2024 are relaxed by Planck PR4 and cosmological supernovae}",
    eprint = "2406.14554",
    archivePrefix = "arXiv",
    primaryClass = "astro-ph.CO",
    year = "2024"
}

@article{T2K:2020,
    author = "Abe, K. and others",
    collaboration = "T2K",
    title = "{Constraint on the matter-antimatter symmetry-violating phase in neutrino oscillations}",
    journal = "Nature",
    volume = "580",
    number = "7803",
    pages = "339--344",
    year = "2020",
    doi = "10.1038/s41586-020-2177-0"
}

@article{NOvA:2022,
    author = "Acero, M. A. and others",
    collaboration = "NOvA",
    title = "{Improved measurement of neutrino oscillation parameters by the NOvA experiment}",
    eprint = "2108.08219",
    archivePrefix = "arXiv",
    primaryClass = "hep-ex",
    journal = "Phys. Rev. D",
    volume = "106",
    number = "3",
    pages = "032004",
    year = "2022",
    doi = "10.1103/PhysRevD.106.032004"
}

@article{Goswami:2026syn,
    author = "Goswami, Srubabati and Gupta, Aman and Rahaman, Ushak and Raut, Sushant K.",
    title = "{Enhancing the sensitivity to neutrino oscillation parameters using synergy between T2K, NO\ensuremath{\nu}A and JUNO}",
    eprint = "2512.00172",
    archivePrefix = "arXiv",
    primaryClass = "hep-ph",
    doi = "10.1007/JHEP06(2026)238",
    journal = "JHEP",
    volume = "06",
    number = "06",
    pages = "238",
    year = "2026"
}

\end{document}